# DESIGNING HUMAN-MACHINE INTERACTIONS IN THE AUTOMATED CITY: METHODOLOGIES, CONSIDERATIONS, PRINCIPLES

**Martin Tomitsch[1,2], Marius Hoggenmueller[1]**
[1] School of Architecture, Design and Planning, The University of Sydney, Sydney, Australia
[2] CAFA Beijing Visual Art Innovation Institute, China

---

## Introduction

Much of human progress is driven by the automation of processes. Automation increases productivity and comfort. The introduction of the conveyor belt into car factories in the 1920s accelerated the process of manufacturing cars – a fundamental contributor to making cars affordable to everyone by decreasing the assembly time per vehicle. The performance and quality of cars has steadily improved over time while keeping the cost for end consumers low, largely due to the ongoing automation of the process through industry robots. Robots come to mind immediately when talking about automation, and robots are becoming more pervasive and increasingly able to complete more complex tasks due to breakthroughs in machine learning and artificial intelligence.

However, automation does not necessarily appear in the form of a physically embodied robot. Early examples of automation in cities are mostly mechanical systems. Urban water networks automated the process of delivering running water to apartments. The sewage system automated the process of removing human waste from one's premises at the push of a button. Traffic lights were introduced to automate the process of regulating vehicle flows at intersections. Even though early traffic lights were technologically not very advanced, they had a significant impact on the way people use the city's infrastructure. New sensor technology led to more sophisticated automation, e.g. using magnetic loops to detect the presence of cars, while also giving people control to manipulate the automated cycle, e.g. through traffic light push buttons for pedestrians. Digital systems further contributed to enabling automation in cities on a much larger scale, such as the automated control of signals on train networks. Software systems are also monitoring the performance of water and sewage networks, creating a closed loop between digital systems and the operation of physical facilities.

As the aforementioned examples demonstrate, technology has always acted as a driver of innovation in cities. Cities and how people live in them have been significantly impacted by technological progress for centuries (Atkinson, 1998). Cars enabled people to move into other parts of their city and commute to work across much longer distances than before (de Waal, 2014, p.62; Mladenović et al., 2019). Access to television meant that people started to socialise within their homes rather than meeting others in the street (de Waal, 2014, pp.37-38). Advances in construction technology combined with the introduction of elevators, which can be considered a form of urban robotic system (Nagenborg, 2018), allowed cities to build taller skyscrapers (Graham, 2014). With the focus shifting from mechanical to digital systems (enabled by new networking and computing technologies), the term “smart city” emerged as a new paradigm to label the digitalisation of cities. A smart city considers the geographical land, citizens, technology and governance to enable environmental sustainability, the creation of smart intellectual capital, citizen participation and the citizens’ well-being (Dameri, 2013). Smart city technologies represent an important driver for the advanced automation of cities and their infrastructure. At the same time, industries are moving from information-communication technology-based approaches towards what has been coined as the fourth industrial revolution, accelerated by emerging technology breakthroughs in fields including automation and robotics (Schwab, 2015). The increasing availability of smart sensors, digital fabrication technologies and artificial intelligence is changing the very fabric of everyday life.

Drawing on these technological advances, the smart city literature explores the role of technology as an enabler for new urban applications across six common domains: economy, environment, governance, living, mobility and people (Camero and Alba, 2019). Particularly relevant within the context of automation are the domains of environment and mobility applications, which includes solutions ranging from smart grids and automated monitoring to transportation and logistics. While much has been written about smart cities from urban studies perspectives (e.g. Townsend, 2013; Kitchn, 2014) and regarding their technological realisation including the application of big data analytics (e.g. Batty, 2013), the opportunities arising from automation and the impact of automation on life in cities have been less explored.

This chapter addresses this gap by unpacking the role of automation in how we experience our cities and use its infrastructure. The chapter introduces design methodologies and assesses their utility for designing the way people interact with the protagonists emerging from this new domain. The chapter takes a human-centred perspective. In other words, the focus here is on citizens and how their daily lives in cities are changed by automation (and what we need to know to shape these changes meaningfully and responsibly). That way, the chapter aims to offer a complementary account of robotics and algorithms in the automation in cities, contributing to the wider discourse on automation in architecture and engineering to change how cities are constructed.

The remainder of the chapter is structured as follows. First, we expand on the notion of smart cities, highlighting research and approaches that are relevant to the automation of cities, and linking this to the field of human-computer interaction. We briefly provide an example of applying a human-computer interaction approach to an automated system in a smart home environment, before introducing a number of methodologies that can be used to structure the process of designing interactions between people and automated urban applications. This is followed by a discussion of specific elements that need to be considered when designing human-machine interactions in an urban environment, demonstrated through specific example applications. The chapter ends with a proposed model for designing automated urban applications (adapted from Tomitsch, 2018) and a set of principles to guide the prototyping and deployment of such applications.

## Smart Cities, Automation and Human-Computer Interaction

There is no uniformly accepted definition of what a “smart city” entails and the term has been used in different ways since its appearance in the literature (Camero and Alba, 2019). Early smart city strategies focused on deploying information technology infrastructure, such as free Wi-Fi (Tomitsch, 2018, p.27), whereas more recently there has been a shift to extend the meaning of “smart” to include sustainability and economic aspects of cities – with technology being seen as the enabler rather than the core of a smart city strategy (Camero and Alba, 2019; Yigitcanlar et al., 2019). City governments are turning to information technology companies, such as IBM, who offer solutions that “help cities of all sizes identify priorities, apply best practices and deploy advanced technologies to help address pressing challenges” (IBM Corporation, 2014). Understandably, such benefits of smart city solutions sound very attractive to governments, who are facing unprecedented challenges due to mass urbanisation and population growth. What these solutions offer is not only a remedy to current challenges but also to future-proof cities by maximising the efficiency of existing infrastructures, from road and rail networks to garbage collection and energy supply chains.

Smart city initiatives are, therefore, closely interlinked with automation in cities. On the one hand, smart city infrastructure is an enabler for the deployment of automated systems. For example, distributed sensors that measure the movement of people and vehicles can be used to ensure the safe passage of automated vehicles through the city. On the other hand, the roll-out of automated systems has an impact on the human-scale interfaces of smart city solutions, such as traffic lights, which may no longer be required in a fully automated future transport scenario (Tachet et al., 2016).

The notion of city apps has been introduced as a way for shifting the focus from the technology elements of smart cities to how they interface with citizens (Tomitsch, 2017; Tomitsch, 2018, p.25). In other words, city apps represent the applications that allow citizens to take advantage of the benefits provided by smart city technologies, reaching beyond efficiency gains by allowing people to make better-informed decisions that improve the way they experience and use city infrastructure. City apps can take the form of physical interfaces (such as traffic light push buttons), digital applications (such as parking smartphone apps) and hybrid solutions (such as touchscreen-based information kiosks). Conceptually, city apps use the city as an operating system, or urban operating platform, with pre-existing input and output mechanisms (Tomitsch, 2018, p.26). Types of input include urban activities, such as traffic or pedestrian flow, and environmental conditions, such as air quality, temperature and ambient light levels. Forms of output include surfaces, such as the street or building façades, and “upgraded” urban furniture, such as smart benches and street lamps.

City apps are defined through six characteristics, which also apply to automated urban applications as a subset of city apps: They are designed to be used by citizens, they are designed to be primarily used in an urban environment, they improve the urban experience for citizens, they make use of digital technology for sensing and visualising data, they can be interactive, reactive or active, and they can be mobile or situated (Tomitsch, 2017). In view of the focus of this chapter, an important consideration is the kind of interaction facilitated between the application and citizens. For example, the interaction between an automated passenger vehicle might be facilitated through a smartphone app that allows a passenger to order a driverless taxi (Owensby et al., 2018) or through the vehicle projecting information onto the road to communicate its internal status to nearby pedestrians (Ngyuen et al., 2019).

The conceptualisation of these forms of interaction builds on knowledge developed through decades of research about the ways people interact with computers, referred to as human-computer interaction (HCI). HCI offers human-centred methods for understanding people's needs and developing interfaces that allow people to interact with technology-based systems. In automated vehicles, such interfaces could take, for example, the form of lights or displays attached to the vehicle. However, the application of human-centred methods for automated systems in urban environments is extremely challenging, as it is associated with high costs of real-world prototypes (e.g. a self-driving car), potential risks of evaluating these systems in the real world and the fact that prospective users cannot yet draw on existing experiences with automated systems (Frison et al., 2017). As a consequence, and also given the infancy of the field, automated urban applications have seen relatively little attention from the field of HCI to date.

However, HCI researchers have begun to investigate interactions between people and automated cars (e.g. Boll et al., 2019; Nguyen et al., 2019) and there is also an established body of HCI literature on automation in smart homes, which offers a foundation for automated urban applications. For example, Jensen et al. (2016) studied a prototype allowing users to set temperature boundaries for their home. The interface controlled an automated system scheduling energy use based on temperature and electricity prices. The study found that some users ended up engaging less with the system, leaving the control to the automated algorithms. While this can be seen as an advantage, the authors point out that lower engagement might also result in missed opportunities for cost savings from not checking the interface regularly. Another key finding from the study, which is echoed commonly in studies of automated systems, was that users expressed uncertainty about how the heat control system actually operated. This effect has been described as the explainability problem and is common to intelligent systems, in particular those that use artificial intelligence algorithms (Došilović et al., 2018). Based on an extensive literature review, Abdul et al. (2018) propose implications and directions for how to build intelligible interfaces into automated systems. An approach cited in the literature is for systems to provide accounts of their behaviours (Dourish, 1995; 1997 – cited in Abdul et al., 2018) and informing users "what [the systems] know, how they know it, and what they are doing with that information" (Belotti and Edwards, 2001).

Research on automation in smart homes, therefore, provides useful knowledge and insights that can be translated to the human-centred automation of cities, as we will further detail in the following sections.

## Methodologies for Designing Citizen-Centred Automated Cities

In addition to being able to provide exemplary studies as a foundation for automating cities, the field of HCI also offers well-documented human-centred methods for how to design automated urban applications. Over time, these methods have been formalised as established methodologies, which typically comprise a collection of methods as well as tools. This section introduces a series of methodologies that are applicable in an urban context when designing interactive urban applications. Each of the methodologies is presented along with examples for how it can be used in the design of automated urban applications. The methodologies included here are: participatory design, action design, design thinking, spiral models, service design and urban design. The methodologies were chosen based on our own practice and experience of conducting research on interactive applications in the urban domain for over 10 years. While it is not necessarily comprehensive, the list provides an entry point in particular for readers, who are less familiar with human-centred design.

The overarching framework that underpins all the methodologies presented here is human-centred design. As the name suggests, human-centred design puts human stakeholders at the centre of the design process. It was initially praised as an alternative to the waterfall model (Royce, 1987), which was a popular model for software development in the 1990s and includes the phases of requirement-gathering, design, implementation, verification and maintenance. The waterfall model received criticism for its sequential nature, which did not allow for iteratively testing proposed solutions with end users and going upstream to repeat one of the earlier phases. As a response, an iterative variation of the waterfall model was introduced, which allowed going up the steps if more information from a previous phase was required. However, it was still not truly iterative and the user was only really considered during the requirements-gathering phase.

The human-centred design model in its simplest form includes the four phases of identifying the needs of human stakeholders, design, prototyping and evaluation (Figure 1). It is truly iterative in that it allows moving back and forward between any of these phases, with the user (ideally) always being at the centre of each phase. There are many different methods and tools that can be used in each of the phases of the human-centred design process. The actual choice of methods and tools depends on the project, its scale, the timeline, the budget, and so on. For example, the needs identification phase can include methods, such as observations, questionnaires, interviews and cultural probes (Tomitsch et al., 2018).

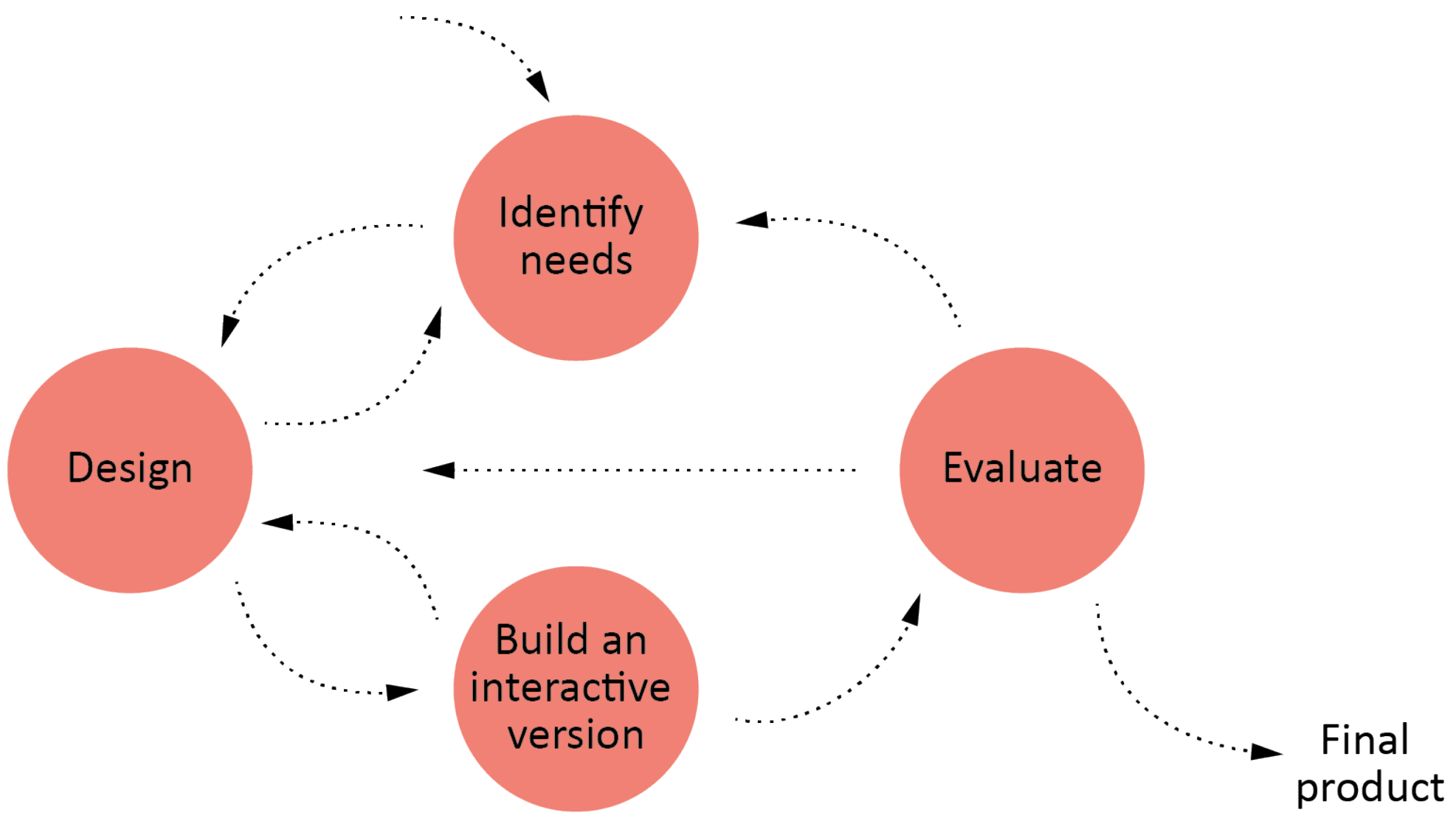

*Figure 1. Simplified human-centred design cycle. (originally published in Making Cities Smarter (Tomitsch, 2018))*

The human-centred design approach provides a useful starting point for automated urban applications. However, its application in an urban context is somewhat complex, as there is not necessarily a clearly defined user group. For example, an automated vehicle needs to consider a variety of human and non-human stakeholders, such as the passengers inside the vehicle, pedestrians around the vehicle, other road users such as motorbike riders and cyclists, and other automated vehicles sharing the same road environment (Boll et al., 2019). In the following, we therefore outline how the various established methodologies within the human-centred design ecosystem can be applied and adapted to the context-specific challenges of automated urban applications.

*Participatory Design*

In participatory design, stakeholders are being elevated to become part of the design team. Participatory design originated in Scandinavia, where it was introduced for the development of software solutions (Floyd et al., 1989). The approach spans the entire design process and as such suggests the use of ethnographic methods, such as field observations and in-situ interviews (Wolcott, 2005) as well as the participatory design of solutions, for example, through sketching and low-fidelity prototyping (Gage and Kolari, 2002).

In a city context, the participatory design approach has similarities with community engagement methods, which have long been used in urban planning for seeking feedback on development proposals and other civic issues. This form of community involvement through consultation is also critical for citizen-centric urban automation, as the success of automated systems, such as driverless shuttles, is highly dependent on user acceptance (Nordhoff et al., 2018). Typically, such consultation is done through online surveys and face-to-face workshops or focus groups. The value and possible direction stemming from such workshops and other methods depend on their structure and the time at which they are held. If held towards the end of a project design phase, they serve more of a formative validation of ideas. A much more effective approach for involving citizens in an urban design process is to run workshops with end users and other stakeholders very early in the project to really allow them to co-create city solutions and to collect input from the local community.

*Action Research and Design Thinking*

Another approach that lends itself well to the design of automated urban applications is action research. The term was first used by MIT professor Kurt Lewin in 1944 to describe a “spiral of steps” in action research (Lewin, 1946). The steps at the time included planning, action and evaluating the result of the action. There are two commonly used types of action research: participatory action research and practical action research. Which of the two is better suited depends on the specific context within which the urban application is designed. Participatory action research is more commonly used for interventions within communities (Paulo, 1970) and has similarities to participatory design in that it suggests involving stakeholders as co-designers. The combination of participatory design and action research methods has also been described to bring advantages for designing applications for urban and neighbourhood community networks, bringing together people, place and technology (Foth, 2006).

The action research phases of planning, acting and evaluating, followed by a reflection, underpin the iterative phases commonly used in design thinking. The practice of design thinking as a creative approach to solving problems first emerged at Stanford University, with IDEO founder David Kelley adopting the use of the term for business purposes. Design thinking is a promising methodology for tackling wicked problems, which are problems that are ill-defined and that cannot be solved by scientific methods alone. Many of the problems that cities are facing today are wicked problems, in that there is no clear solution, and solutions might lead to new, unforeseen problems. By using the iterative phases of design thinking (Figure 2), one is able to handle the complexity of wicked problems in the urban environment.

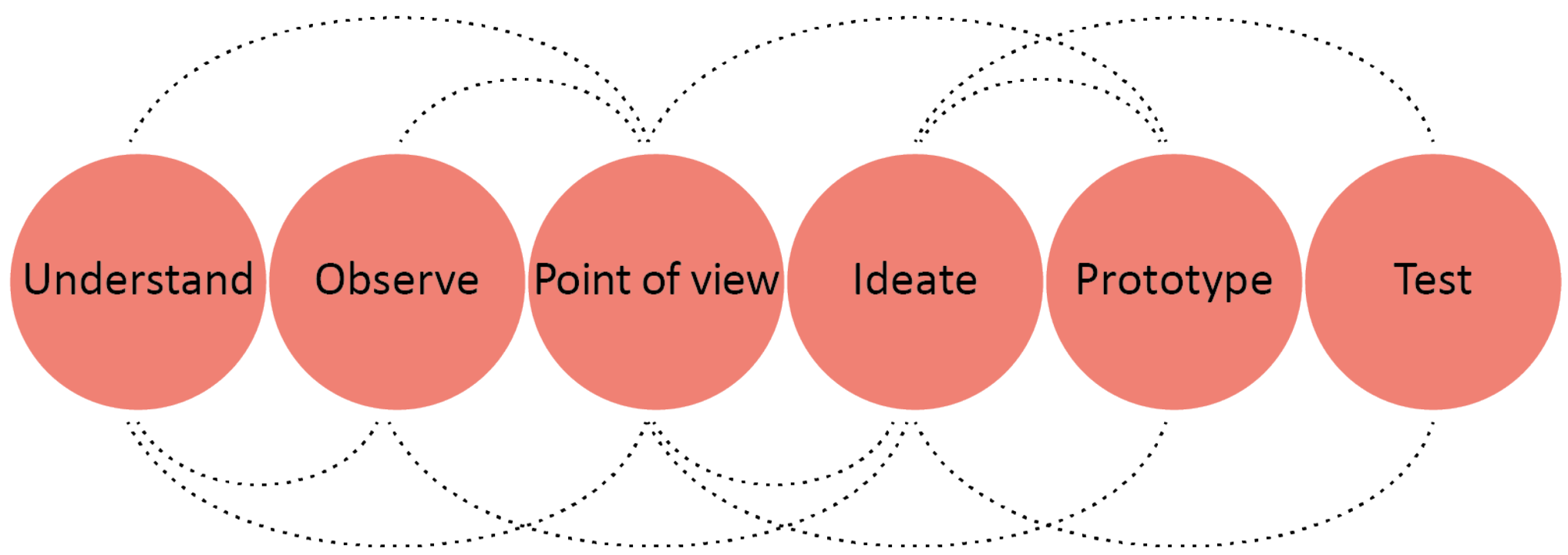

*Figure 2. Phases of a typical design thinking model. (based on Stanford University’s original design thinking model) (adapted diagram originally published in Making Cities Smarter (Tomitsch, 2018))*

The design thinking model is useful within the context of the automated city as it prescribes the grounding of the design solution in the problems observed in the real world. Following this process counteracts the risk of introducing new applications based on the availability of a new technology. Such an approach can be problematic as technology-centric solutions are rarely accepted by end users. The prototyping step in the design thinking model allows for testing ideas early. The goal here is to prototype concepts quickly and often, rather than creating a fully functional solution, which may fail to actually address any real needs in the city. While a plethora of prototyping methods have been documented (cf. Tomitsch et al. (2018) for an overview of prototyping methods), this step represents unique challenges when prototyping automated urban applications. For example, prototyping an automated vehicle is technically complex and costly. During the early stage of a design process, some elements can be prototyped and tested within a controlled environment, for example, by using the experience prototyping method (Tomitsch et al., 2018, p.58). Another approach that we have used in our own work is the use of hyperreal prototypes (Hoggenmueller and Tomitsch, 2019). The prototypes use 360-degree video recordings of a real urban situation (e.g. an approaching automated vehicle), which are then imported into VR and extended through interactive functionality, to increase a sense of realism in users. Hyperreal prototyping, therefore, allow the testing of complex automated systems with potential users within a low-risk environment. The findings generated from the iterative process of prototyping and testing are then fed into the technical development of the automated urban application.

*Spiral Models*

Spiral models encompass the idea of incrementally developing and exploring an idea. Two commonly used spiral models in HCI are Verplank's spiral (Verplank, 2009) and Resnick's creative learning spiral (Resnick, 2007). Verplank's spiral (Figure 3, left), also known as 'hunch-and-hack' approach (Matias, 2012) starts with the designer and the designer's idea (a hunch) that leads to an initial representation in the form of a hack. The hack allows the more detailed elaboration of an idea, resulting in a prototype, which helps the designer to understand and formulate principles. These principles go beyond the prototype in that they set out the guidelines for products, which can in some cases even lead to a new paradigm and thus open a new market opportunity. This way of thinking can be beneficial for designing automated urban applications, as it encourages going beyond a specific product and exploring potentially new paradigms. For example, the design of automated food delivery robots might lead to a new paradigm for urban food consumption.

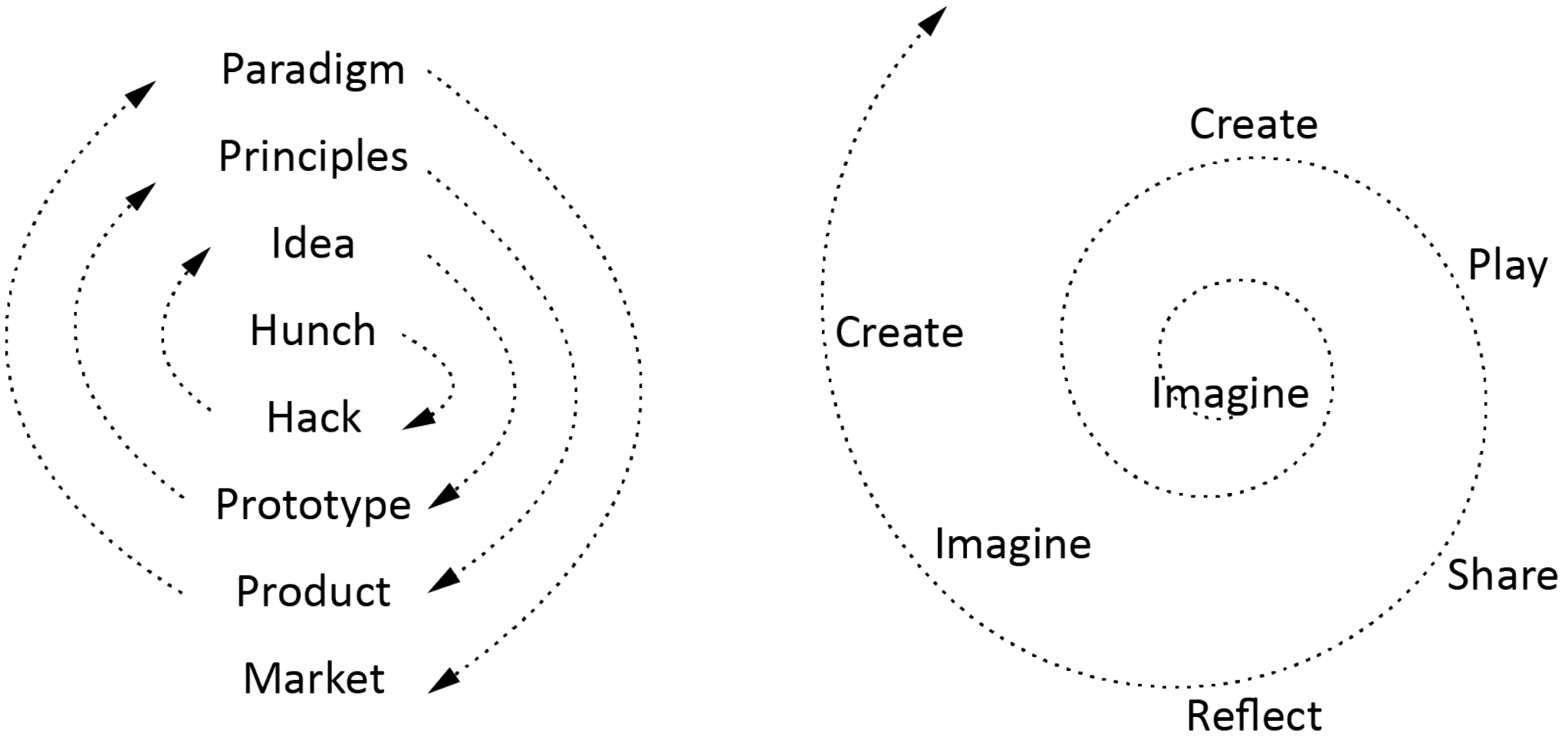

*Figure 3. Two models for designing interventions: Verplank's spiral (left) and Resnick's spiral (right). (originally published in Making Cities Smarter (Tomitsch, 2018))*

Resnick's creative learning spiral also puts the designer at the centre, but is deeply iterative and never ending (Resnick, 2007) (Figure 3, right). It shows some similarities to the phases in design thinking, in that it suggests a process that involves creating and reflecting, before moving on to the next iteration cycle.

*Service Design*

As a methodology, service design provides a framework for designing services with the aim to improve their quality as well as the interactions between the service providers and their customers. Service design has its roots in the marketing and managing disciplines, where it was introduced to describe the importance of the integrated design of material components (products) as well as immaterial components (services) (Shostack, 1982). Service design suggests a shift from focusing on products towards focusing on services. The service design methodology incorporates the principles of being user-centred, involving a co-creative process, breaking services into sequences, making services tangible through evidencing and striving for a holistic viewpoint (Stickdorn et al., 2011).

People experience their city to a large extent through its services. Taking public transport, using a bicycle lane or sitting down on a park bench – these are all services provided by the city. To help identify what elements are the constituents of a service, the service design methodology uses the notion of touch points. Touch points can be thought of as the interfaces through which a customer (or citizen) experiences a service. Identifying touch points in the city and analysing their quality can be an effective first step when designing automated urban applications. At the same time, it is essential to keep a focus on the network of touch points to understand how changing one touch point affects the holistic experience. Many of the current and speculative examples of automated systems in cities represent some type of services, whether it is the automated supply of electricity, the automated regulation of indoor environmental quality in buildings, the automated delivery of pizzas or the automated collection of rubbish bins. Applications, such as automated vehicles, are also expected to have an impact on existing services, which may raise ethical considerations such as equitable access to services (Mladenović et al., 2019).

*Systems Thinking*

The systems thinking methodology provides a framework for understanding how individual elements in a network of elements influence each other. Early concepts of system thinking were formulated and applied for organisational learning, but today it is used as a problem-solving approach across many disciplines including urban planning. Two commonly used approaches to systems thinking are hard systems and soft systems methodologies. Hard systems are quantifiable and use computer-based algorithmic methods such as simulations to tackle problems. This approach works well for tackling relatively simple problems but is less suited when dealing with systems that depend heavily on human parameters since it is "unable to deal with multiple perceptions of reality or with extreme complexity" (Jackson, 1993). Simulations and other quantifiable methods are therefore only of limited value for the design of automated urban applications, which are built around humans. In contrast, the soft systems methodology is better suited for solving problems in messy situations that involve people holding multiple and conflicting frames of reference (Checkland and Poulter, 2010).

Systems thinking can be used to identify the root causes of issues that cities are dealing with and to better understand the holistic value of changes to the city as a system, and how to implement such changes through tactical strategies (Dirks et al., 2010, p14.). In other words, this allows a more strategic implementation of automated systems, guided by the aim to address root causes through the deployment of an automated solution (Macrorie et al., 2020). Such an approach is also useful to prevent the pursuit of technology-driven. As Price proposed, "Technology is the answer. But what is the question?" (Pierce, 2006). Hill (2015) suggests to flip the question to "reframe the way we think about today's cities, as well as those tomorrow. … The answer 'autonomous vehicles' suggests the question 'what do we want our streets to be like?'" This proposition is critical to ensure that cities will continue to develop as livable environments rather than being driven by technological opportunities.

*Urban Design*

Urban design is more of a discipline than a methodology, but it is relevant in the context of this chapter as it offers a range of principles for designing the urban, public realm. As a framework, these principles can provide critical guidance for the design of smart cities, and in extension, for automated urban applications. For example, urban design principles comprise the character of the urban space, continuity and enclosure, quality of the public realm, ease of movement, legibility, adaptability, accessibility and diversity (Fisher, n.d.). Automated urban applications may be designed so that they improve certain aspects of those elements. For example, accessibility could be improved by providing automated shuttles that transport people with physical disabilities between their home and the nearest train station.

At the very minimum, it is necessary to ensure that automated interventions do not break any of these principles. For example, it is important that designers of automated urban applications, such as delivery robots, understand the flow of people and ensure that their intervention does not negatively affect how people move through the space. Many government organisations have developed and maintain their own set of principles for urban design, such as the *Urban Design and Public Realm Guidelines* developed for a precinct in Sydney, Australia (Woods Bagot, 2012).

## From Research Trials to Speculative Prototypes: Design Considerations in Automated Urban Applications

To date, the number of automated systems that interact directly with citizens is still extremely limited. In many cases, such as the automated trains used at airports and increasingly in cities, the automation is implemented in a way that is almost unnoticeable to the end user. Unless passengers deliberately look for a human driver, they may not even be aware that their service is operated by algorithms. In many cases, there are still human operators in charge or at least monitoring the service, but doing so remotely from a centralised control room. These types of early automation are indeed examples for successful implementations from a user-experience perspective, as they hide the technology almost entirely from the end user. As such, the reality of automated transport is fundamentally different to some of the visions portrayed in science fiction movies from the last century. – There are no humanoid robots with a dark sense of humour, like Johnny Taxi in the 1990 fantasy movie *Total Recall*, operating the driverless transport systems of our time. In this section we illustrate through examples how the design approaches outlined above can be used to determine particular elements of automated systems in cities.

### *Communicating Intent and Awareness*

As automation is becoming more advanced and widespread, its effect on citizens will become increasingly more pronounced. Trains that are automated may not directly affect passengers and others; however, automated vehicles roaming the streets pose significant challenges as they need to be designed to be able to interact with pedestrians and other road users. Here, the challenge is not of purely technical nature in terms of how to engineer cars so that they drive themselves along a given path. Introducing automated vehicles into the streets of cities also poses a wide range of socio-technical challenges related to the human factors of these new vehicles. Beyond the issue of societal acceptance (Nordhoff et al., 2018), automated vehicles need to be able to deal with the unpredictable behaviour of pedestrians and other human road users, who will have to share the roads with automated vehicles. An approach that uses HCI principles is to communicate a vehicle’s intent and awareness through displays or other digital features on the outside of the vehicle (Rasouli and Tsotsos, 2019). Solutions include using low-resolution (low-res) displays (Mercedes-Benz, n.d.), displays attached to the vehicle (Clamann et al., 2017), adjustment of existing features (such as the vehicle’s headlights (Chang et al., 2017)) and projections onto the road in front of the vehicle (Nguyen et al., 2019).

In the context of a larger project, carried out between researchers from robotic engineering, urban planning, social science and interaction design, we are currently investigating how these socio-technical challenges can be addressed through an HCI-led design process, thereby applying and tailoring some of the previously introduced methodologies. Our specific case study deals with the design of a low-res lighting display that could be attached to any automated vehicle to communicate its intent and awareness to surrounding pedestrians and other road users. A low-res lighting display was chosen as the enabling technology due to its high contrast and its ability to communicate information through simple visual cues (i.e. colours, temporal change of visual elements) which are also "readable" from a distance (Hoggenmueller and Wiethoff, 2015), and even on moving objects. Further, low-res displays based on light-emitting diodes (LEDs) are highly flexible in terms of size, resolution, shape and integration into various product designs (Hoggenmueller et al., 2018), which makes it more likely for them to become applied by car manufacturers when compared to conventional screens.

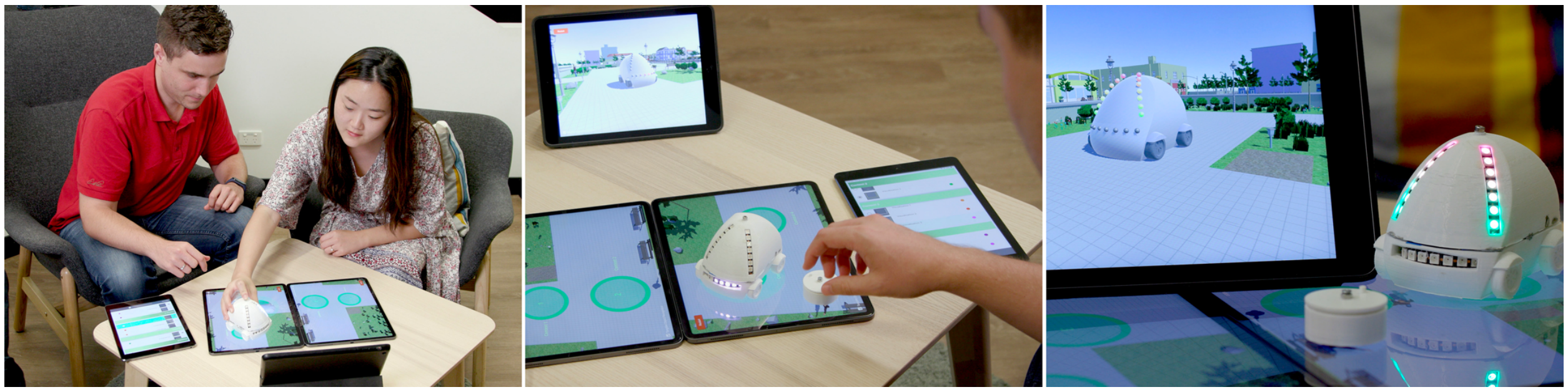

*Figure 4. Involving end users and other stakeholders in co-creation design focus groups. The physical model of the automated vehicle enables participants to experiment with and evaluate different light patterns for communicating the vehicle's intent and awareness in urban situations. The iPads offer dynamic bird's-eye and 3D views of the urban environment.*

Following some of the principles from participatory design, such as involvement of stakeholders early in the design process and co-creation of solutions, we started our investigation with a series of focus groups. To make such a highly fictitious scenario more graspable and therefore inclusive (Malizia et al., 2018), we developed a prototyping toolkit, consisting of an iPad application and a miniature 3D-printed model of the automated vehicle and the embedded low-res lighting display. Supported through this toolkit, participants were advised to "play through" a variety of potential scenarios that could occur in urban spaces with automated vehicles and adjust the light patterns on the miniature prototype (Figure 4). Moving from low-fidelity prototyping to a higher-fidelity representation, we designed a final set of light patterns based on the feedback from the design focus groups. While the final light patterns were implemented on an actual automated vehicle – developed by our collaborators from robotic engineering – we could not run user studies in the real world due to safety concerns. We therefore used our approach of hyperreal prototypes (Hoggenmueller and Tomitsch, 2019), for which we videorecorded the scene with a stereoscopic 360-degree camera and then imported the immersive videos into a virtual reality (VR) environment to be experienced by participants via a VR headset (see Figure 5). This allowed us to evaluate high-risk situations with automated vehicles, in a way which is time-efficient and scalable also for comprehensive user studies with a large number of participants. At the same time, the approach enables a more realistic representation of the urban environment compared to computer-generated simulations.

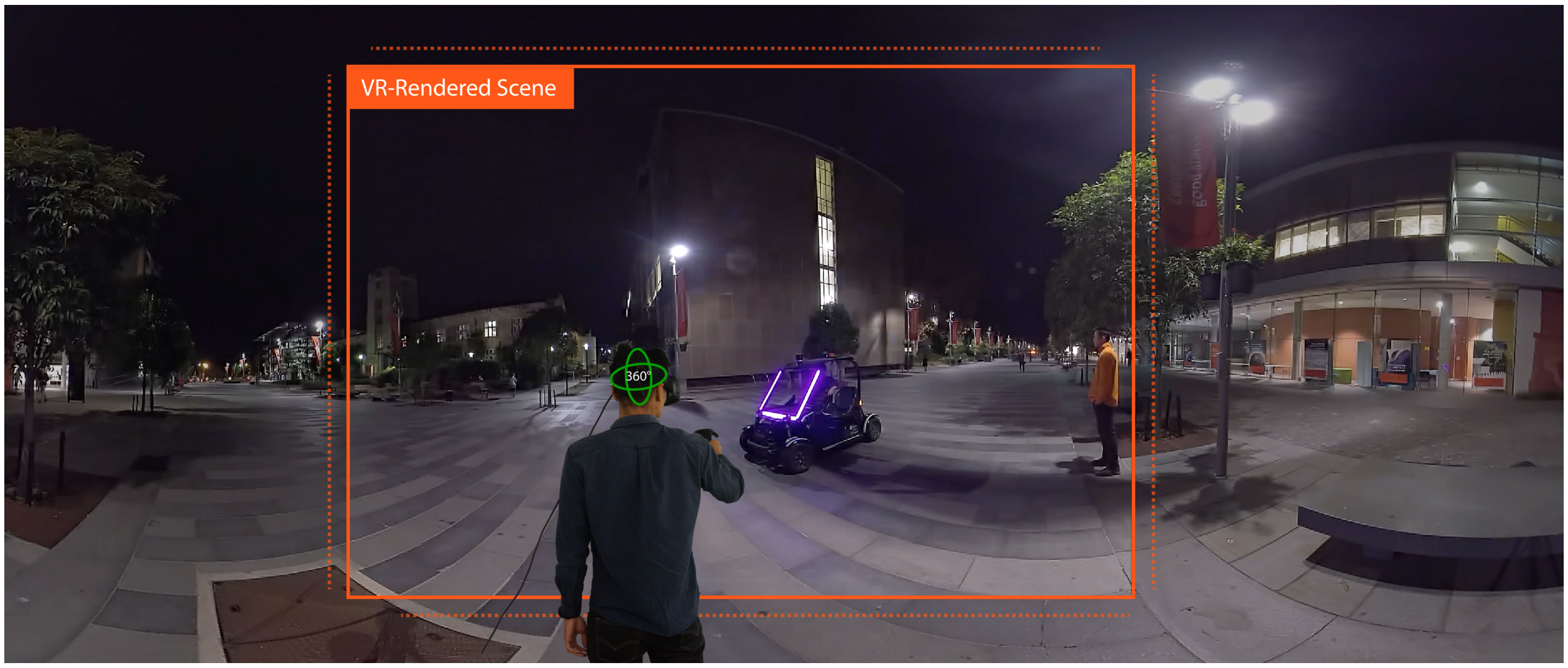

*Figure 5. Evaluating interactions with automated vehicles in VR: using a 360-degree video recording of an actual autonomous vehicle equipped with a low-resolution lighting display, which is imported into Unity3D so that study participants can experience the situation through a VR headset.*

*Determining the Physical Representation of Automated Systems*

In some cases, it may be desired or even necessary to carefully design a physical representation for an automated system – with the purpose to guide a human interaction and/or to give the system a distinctive character. These considerations are usually relevant for embodied automated systems. An example for such a system, is the wayfinding communication robot that has been trialled at a train station in Kyoto in Japan (Figure 6, left). The researchers behind this trial used a humanoid robot to draw the attention of passers-by (Shiomo et al., 2008). Another example is an autonomous city explorer robot, which was deployed in the city centre of Munich (Figure 6, right) (Bauer et al., 2009). The robot was designed to navigate through the urban environment just by interacting with pedestrians, instead of relying on the use of GPS technology. The aim of the project was to investigate a natural and intuitive human-robot interaction in public spaces, thereby reversing the usual scenario in which a robot serves a human. These are therefore conceptually different applications of automation, compared to, for example, driverless trains, where the train network represents the robotic system driven by software algorithms. The two examples described above show that in some cases it might be indeed desired to create a human-like embodiment of an automated system.

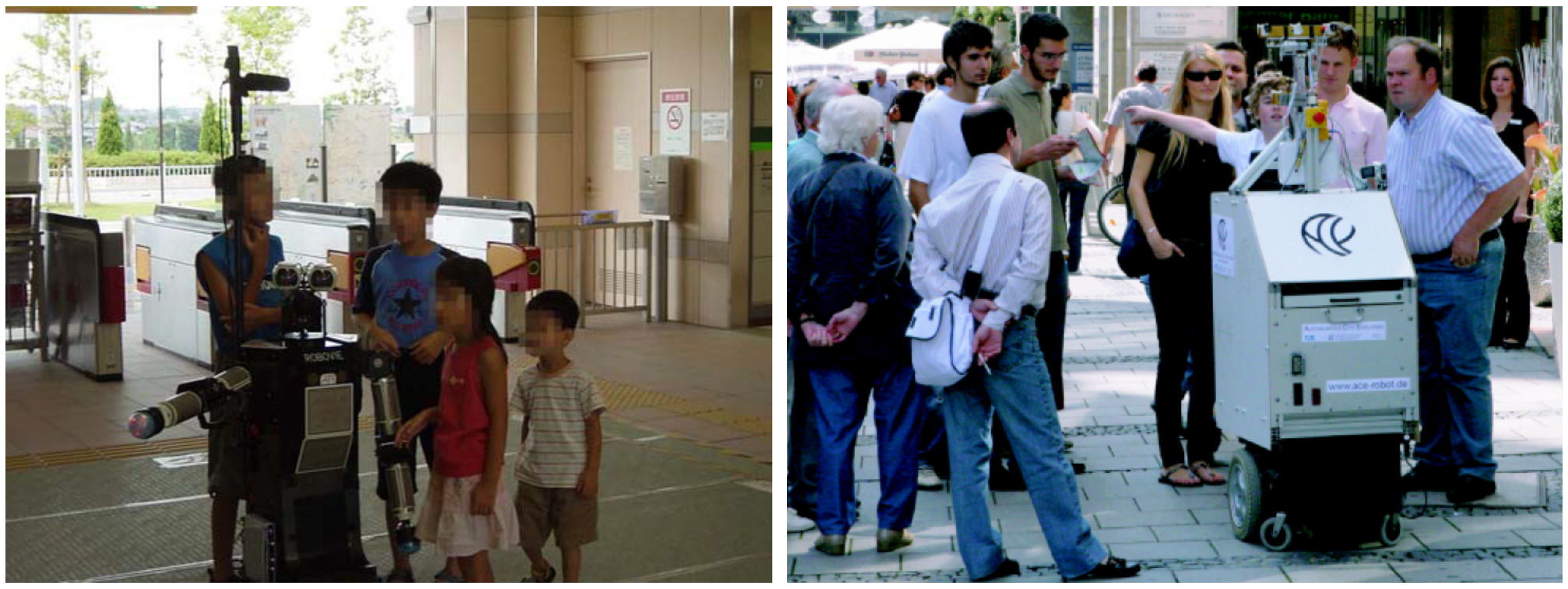

*Figure 6. Field trials of a wayfinding communication robot at a train station in Kyoto, Japan (left; image source: Shiomo et al., 2008) and an autonomous city explorer robot in Munich, Germany (right; image source: Bauer et al., 2009).*

Research studies in HCI have also begun to investigate what effect the representation of a robot has on the emotional response from passers-by (DiSalvo et al., 2002; Luria et al., 2018). Beyond triggering an emotional response from passers-by, giving an automated application a character can help to visualise its purpose. To that end, the Poop Bot featured in the 2018 Netflix series *Maniac* has a clear design affordance that communicates its purpose of being: to collect biological dog waste. Its animated lights and sound effects give it an almost human character, and help to convey its struggles and sad existence. This embodiment creates a sense of feeling sorry for the bot, successfully overwriting other emotional responses that might be associated with the mental image of picking up dog waste.

*Enabling Direct Operation Through an Interface*

Interfaces in HCI allow users to interact with a computing system. An interface contains the elements needed for a user to trigger specific commands and to receive feedback about the status and outcome of an operation. In smart city applications, the interface commonly takes the form of operation dashboards (for example used by the city government to control a train network), mobile applications (for example used by citizens to record broken infrastructure or to unlock a bicycle), and in some cases physical control elements (such as the button on pedestrian traffic light systems). The interface allows the user to trigger some form of operation that has an effect on the underlying software system. In urban automated applications, this means that the interface allows the user, which may be a dedicated operator or a passer-by depending on the application) to control the automated process carried out through the application.

For example, a mobile application would be used in the fictional scenario of ordering a pizza delivered by an automated delivery robot. The application would further provide the customer with feedback about the status and whereabouts of the robot. Much of this interaction flow already exists for conventional pizza delivery systems. The challenge will be to adapt those existing interfaces to cater for automated systems.

Automated vehicles could feature sensors that allow pedestrians to control their behaviour to a certain extent. For example, a person that has ordered a driverless taxi might wave at the vehicle to confirm their location – a feature that has been identified in research studies as desired by prospective users (Owensby et al., 2018; Nguyen et al., 2019). The interaction between pedestrians and automated vehicles (or other urban robots) can potentially be modelled based on gestures used by traffic control officers, which a study found to include arm movement, eye behaviours, body postures and head movements (Gupta et al, 2016). Another promising approach, which can borrow from smart home assistants, is the use of voice-based commands – an approach that has been trialled, for example, in the above-mentioned study of an autonomous city explorer robot asking passers-by for directions (Bauer et al., 2009; Weiss et al., 2015).

The specific implementation of the three elements discussed in this section – communicating intent and awareness, determining the physical representation, and enabling direct operation – depends on the particular application and its context. Using the design methodologies outlined above in combination with prototyping and deployment studies can help determine the best approach for a given situation. One of the challenges for designing automated systems in an urban environment, is the observation that urban interventions are highly dependent on physical context, the demographic of the target audience, their cultural norms and expectations, and so on. Following a human-centred design approach can help to address this challenge through involving end users and consulting relevant stakeholders during the design process.

## Principles for Prototyping and Deploying Automation

Similar to city apps (Tomitsch, 2017; Tomitsch, 2018, p.25), the design of automated urban applications is never-ending (Figure 7). Digitally enabled applications have a different lifespan as compared to the physical infrastructure (Tomitsch, 2018; p.208) – both in terms of technology becoming obsolete and superseded as well as in terms of people's expectations driven by the rapid advancement of consumer electronics. Similar to other road infrastructure, automated urban applications need to be maintained, which also opens up the need (and opportunities) for conceiving new business models. For example, bus stops in some cities are being maintained through an advertising provider, who in return are granted permission from the city to use the bus stops for displaying their advertising.

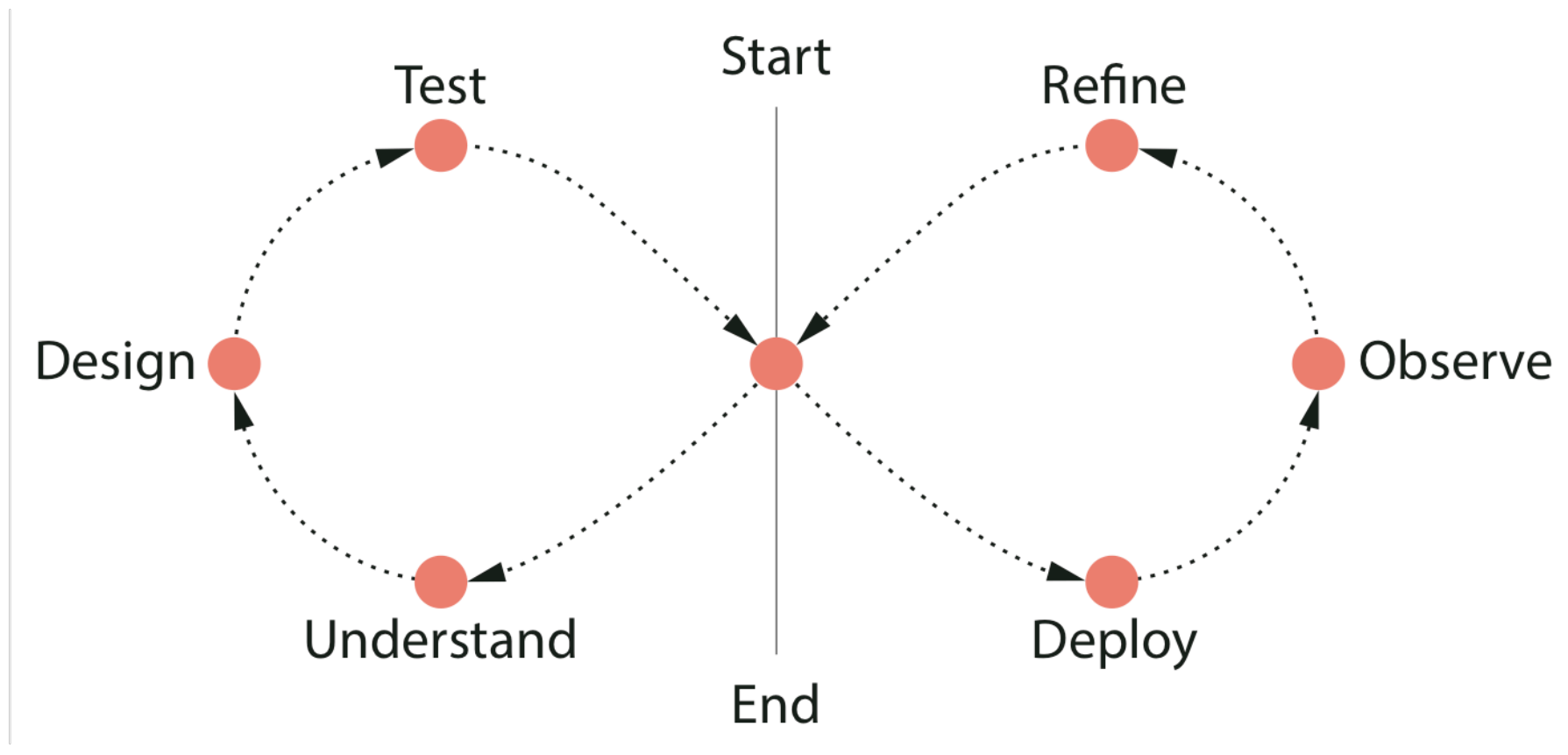

*Figure 7: A model for designing urban applications. (Originally published in Making Cities Smarter (Tomitsch, 2018))*

Based on the examples discussed throughout this chapter and our own work in the field of smart cities and urban robots (Tomitsch, 2017; Tomitsch, 2018; Hoggenmueller et al., 2018, Hoggenmueller and Hespanhol, 2020), we propose a series of design principles (adapted from Tomitsch, 2018) and illustrate how they can be applied in the context of automated cities. The principles specifically focus on the prototyping and deployment of automated urban applications, and complement the design considerations outlined in the previous section. Table 1 illustrates how the principles are linked to the design methodologies presented earlier in this chapter. The table captures which methodology to draw on for selecting methods and tools when implementing a specific principle.

| Principles \ Methodologies | Participatory Design | Action Research | Spiral Models | Service Design | Systems Thinking | Urban Design |
|---|---|---|---|---|---|---|
| 1) Starting with simple prototype representations | | ● | ● | | | |
| 2) Exploring interface and experience aspects simultaneously | | | | ● | | ● |
| 3) Prototyping for testing | | ● | | | | |
| 4) Testing and iterating instead of doing trials | ● | ● | ● | | ● | |
| 5) Preparing prototypes for deployment | | | ● | | ● | ● |
| 6) Launching with the citizen in mind | ● | | | ● | | ● |
| 7) Planning for failure and maintenance | | | | ● | ● | |
| 8) Knowing the regulations and how to negotiate them | ● | | | | ● | ● |
| 9) Managing expectations and biases | ● | | | ● | | |

*Table 1. Principles for prototyping and deploying automated urban applications, and their mapping to design methodologies for achieving citizen-centric solutions.*

The principles offer guidance for designing automated urban applications. However, they are not meant to necessarily be comprehensive, and in some cases, it might also be justified and necessary to violate or adapt some of the principles. Though the principles are underpinned by a human-centred perspective, they can also offer useful starting points for automation that is invisible to the end user, such as the example of driverless shuttle trains.

***Starting with simple prototype representations.*** Creating simple, quick prototype representations (e.g.in the form of a mock-up or as an experience prototype (Tomitsch et al., 2018)) allows thinking through, communicating and testing aspects of the envisioned automated urban application. Early prototypes can be speculative to evoke “big picture” conversations or representations of detailed aspects to test specific elements of an envisioned automated urban application. Speculative, simple prototype representations can also facilitate discussions with stakeholders to test the feasibility of an idea and to identify potential political or legal challenges (Tompson and Tomitsch, 2016). Previous work as also highlighted the importance of understanding “prototypes as filters” in order to focus on different qualities of interest in each design iteration (Lim et al., 2008). This is in particular relevant for automated urban applications, for which a high level of complexity in early instantiations might distract designers and stakeholders from the important design features that are under investigation.

***Exploring interface and experience aspects simultaneously.*** The interface of an automated urban application is composed of input and output controls, which form the interface through which people can operate the application. In many ways, it is easier to prototype the interface of an application compared to the experience. However, it is important to keep both in mind throughout the prototyping process as ultimately the prototype will be used in an urban context, which shapes and affects the experience of end users. Prototyping the experience along with creating interface prototypes ensures that the application is built with the urban context and its unique characteristics in mind. Using immersive and realistic representations of the urban context – for example, in the form of hyperreal prototypes – can be a potential measure to explore interface and experience aspects simultaneously to a certain extent.

***Prototyping for testing.*** The development of prototypes should be informed by research questions, which also determine the test method. Prototypes can be tested in lab or “in-the-wild” environments. Specific prototype representations are more suited to one or the other approach, although the fidelity of a prototype is not necessarily linked to whether it should be tested in the lab or in the wild. For example, the hyperreal prototypes discussed in this chapter offer a high level of detail and realism but are intended to be tested in a controlled lab environment.

***Testing and iterating instead of doing trials.*** Trials are a common approach to testing smart city initiatives, commonly following a fail-or-work mentality (Tomitsch, 2018, p.188). Instead of thinking in trials, new initiatives should ideally be tested through prototypes, encouraging iteration and incremental improvements. If this is difficult to achieve because of the complexity of the urban environment, an alternative approach is doing tests in surrogate environments, such as university campuses or urban festivals. For example, the automated vehicle study presented in this chapter was carried out on our university campus as a proxy for a shared environment. Positioning the study as an iteration in a larger project allowed us to develop and test particular elements of the automated system. Using hyperreal prototypes further reduced any potential risk to study participants as they were able to experience the automated vehicles in VR. While our interdisciplinary project team is simultaneously exploring opportunities for trials with a public transport provider, these trials require much longer timelines and more complex approval processes.

***Preparing prototypes for deployment.*** During the development of prototypes, it is critical to already consider their subsequent deployment. Specifics of the urban environment, such as the physical location, the people in the location and even urban wildlife, may pose requirements that need to be addressed in the design of the prototype. This includes the fail-testing of prototypes within controlled environments to identify unforeseen human behaviour leading to failure in the system. Here again, the use of surrogate environments or hyperreal prototypes offer a low-risk approach for fail-testing urban automated applications with potential end users.

***Launching with the citizen in mind.*** When changing existing urban environments, infrastructures or services, it is also necessary to carefully manage change and consider the impact on the local community. To successfully launch automated urban applications, citizens need to be seen as active contributors to city life. This means their role and potential participation in the deployment needs to be considered beyond simply informing them about initiatives. For example, the long-term success of dock-less bicycle systems is dependent on involving not only local governments but also citizens through partnerships (Li et al., 2018). Participatory design can provide a framework for forming these partnerships, leading to the development of sustainable business models. Automated urban applications, such as shared automated shuttles, are likely to face similar challenges regarding their long-term acceptance and uptake.

***Planning for failure and maintenance.*** The more complex the underlying technology, the more likely urban automated applications are to fail. It is, therefore, important that the deployment of automated urban applications includes strategies for dealing with failure. This can include automated scripts that restart the software driving the application and sending messages to off-site maintenance teams. The design process of a city app does not end with its deployment, making it essential to also consider update cycles and provisions for regular maintenance. Automated systems can also benefit from lessons learned in public display deployments, offering suggestions for better integrating failure and maintenance strategies into the design process (Parker et al., 2018). Such strategies are also essential for convincing industry stakeholders in the investment of these costly deployments (Hosio et al., 2014).

***Knowing the regulations and how to negotiate them.*** Deployments in public space require approvals from relevant authorities and need to follow local regulations that will be specific to the deployment location. Different regulations apply to different parts of the city, such as intersections or busy main roads versus pedestrian zones or residential neighbourhoods. Established, local urban design principles – introduced as one of the design methodologies in this section – can be helpful in understanding those local requirements. Following the ideology of participatory design and involving city authorities as stakeholders early-on in the design process can further be an effective strategy for bringing those stakeholders that maintain regulations along, enabling them to inform the project from the inside rather than imposing regulations from the outside.

***Managing expectations and biases.*** In order to ensure their successful adoption, it is important to manage the expectations of all stakeholders, who have an invested interest in an automated urban application. For example, this can be done through staged deployments or through evaluating and revising interventions. Sensitive deployments should ideally be preceded by low-profile prototype trials, which can be achieved, for example, by collaborating with universities or other research institutions. In particular, the introduction of automated systems is likely to lead to mixed expectations as citizens do not yet have experience in interacting with such systems in an urban context. People might be biased by current transportation paradigms, leading to presumptions that hinder to traverse the full spectrum that these applications offer. It is therefore critical to be aware of and address these biases during the design process already. In our automated vehicles study, for example, we attempted to lower these biases during the focus group sessions by using a very simplistic physical model of a car (Figure 4), with the low-res lighting display being the only fully implemented feature.

## Conclusion

As we outlined in this chapter, interactions between citizens and automated city infrastructure do not necessarily involve people interacting with humanoid robots. Examples such as automated vehicles push the notion of the city itself being a distributed robot (Jacob, n.d.; Hill, 2015). This enables a broader perspective on opportunities for automating cities and their infrastructure, from moving people or goods to emptying bins and cleaning the façades of skyscrapers.

Thinking about the city's infrastructure as a robot also helps to counteract the risk of the excitement around automation encouraging technology-centred thinking. Instead, the opportunities arising from automation can be seen as prompts for creating more liveable cities, whether automated or not (Hill, 2015). At the same time, when automating existing services, it is important to consider not only the impact on people's existing experiences but also the impact on the built environment. For example, automating services such as the purchasing of public transport tickets leaves behind unused, outdated infrastructure as ticket counters are no longer needed. At the same time, this creates opportunities for more broadly rethinking what new roles the obsolete infrastructure and space can take on. For example, ticket counters could make space for a cafeteria or to offer other services that are relevant to passers-by in a train station environment. The High Line in New York represents another well-known example for successfully repurposing out-dated city infrastructure in a way that improves urban life.

Many challenges remain, some of which are outlined in other chapters of this book, such as the appropriate involvement of communities in the design of cities and ethical considerations linked to the collection of data and the use of advanced sensors that help urban robots navigate the streets with the side effect of recording images of passers-by. It is also timely to not only consider how to design the automated city as an ethical and equitable city but to also investigate how automation can address complex challenges such as economic transformation and climate change. The methodologies, considerations and principles introduced in this chapter contribute to establishing a foundation that has the ability to serve as a framework for guiding future research and deployments of automated cities.

**References and Chapter Bibliography**